\documentclass[12pt]{article}
\usepackage[OT1]{fontenc}
\usepackage[utf8]{inputenc}
\usepackage{tocloft}
\usepackage{amsfonts}
\usepackage{amsmath}
\usepackage{amssymb}
\usepackage{array}
\usepackage{slashed}
\usepackage{bigints}
\usepackage{pifont}
\usepackage{bm}
\usepackage{bbm}
\usepackage{upgreek}
\usepackage{booktabs}
\usepackage[nosort]{cite}
\usepackage{xcolor}
\usepackage{dsfont}
\usepackage{float}
\usepackage{framed}
\usepackage{graphicx}
\usepackage{indentfirst}
\usepackage{mathrsfs}
\usepackage{multirow}
\usepackage{pdflscape}
\usepackage{setspace}
\usepackage{subdepth}
\usepackage{subfig}
\usepackage{titlesec}
\usepackage{wrapfig}
\usepackage[all]{xy}
\usepackage{young}
\usepackage[vcentermath]{youngtab}
\usepackage{relsize}
\usepackage{stackengine}
\usepackage{verbatim}
\usepackage{slashed}

\usepackage{hyperref}
\hypersetup{colorlinks=true}
\hypersetup{linkcolor=black}
\hypersetup{citecolor=black}
\hypersetup{urlcolor=black}

\numberwithin{equation}{section}

\usepackage[left=2.5cm,right=2.5cm,top=2.5cm,bottom=3cm]{geometry}
\fontdimen2\font=1.2\fontdimen2{\jot}{5pt}

\newlength{\bibitemsep}
\newlength{\bibparskip}
\let\oldthebibliography\thebibliography
\renewcommand\thebibliography[1]{%
  \oldthebibliography{#1}%
  \setlength{\parskip}{\bibitemsep}%
  \setlength{\itemsep}{\bibparskip}%
}

\titleformat{\section}{\bfseries}{\thesection.}{4pt}{}
\titlespacing{\section}{0pt}{20pt}{6pt}

\titleformat{\subsection}{\normalfont\itshape}{\thesubsection.}{4pt}{}
\titlespacing{\subsection}{0pt}{15pt}{6pt}

\titleformat{\subsubsection}{\normalfont\itshape}{\thesubsubsection.}{4pt}{}
\titlespacing{\subsubsection}{0pt}{15pt}{6pt}

\titleformat{\paragraph}{\normalfont\itshape}{\theparagraph.}{4pt}{}
\titlespacing{\paragraph}{0pt}{15pt}{6pt}

\renewcommand{\tilde}{\widetilde}
\renewcommand{\t}{\tilde}
\renewcommand{\hat}{\widehat}

\renewcommand{\bar}{\overline}

\newcommand{\half}{\frac{1}{2}}

\newcommand\ep{\varepsilon}

\DeclareMathOperator{\tr}{tr}
\DeclareMathOperator{\Tr}{Tr}

\DeclareMathAlphabet{\mathbfsf}{OT1}{cmss}{bx}{n}

\newcommand{\alphadot}{\dot\alpha}

\newcommand{\TT}{\mathsf{T}}

\newcommand{\SL}{\mathscr{L}}

\newcommand{\bR}{\mathbb{R}}
\newcommand{\bZ}{\mathbb{Z}}

\newcommand{\cA}{\mathcal{A}}
\newcommand{\cB}{\mathcal{B}}

\newcommand{\cF}{\mathcal{F}}

\newcommand{\cH}{\mathcal{H}}
\newcommand{\cI}{\mathcal{I}}

\newcommand{\cL}{\mathcal{L}}
\newcommand{\cM}{\mathcal M}
\newcommand{\cN}{\mathcal{N}}
\newcommand{\cO}{\mathcal{O}}

\newcommand{\cQ}{\mathcal Q}
\newcommand{\cR}{\mathcal{R}}
\newcommand{\cT}{\mathcal{T}}

\newcommand{\cW}{\mathcal{W}}

\newcommand{\cY}{\mathcal{Y}}

\newcommand{\ed}{\,.}
\newcommand{\ec}{\,,}
\newcommand{\QFT}{\text{QFT}}

\newcommand{\be}{\begin{equation}}
\newcommand{\ee}{\end{equation}}
\newcommand{\beq}{\begin{equation}}
\newcommand{\eeq}{\end{equation}}

\newcommand{\one}{^{(1)}}

\allowdisplaybreaks 

\DeclareFontShape{OT1}{cmr}{mx}{n}%
{<->cmr10}{}
\newcommand{\mytitlefont}{\fontseries{mx}\selectfont}
\DeclareMathAlphabet{\titlemath}{OT1}{cmr}{mx}{n}

\DeclareMathOperator{\CS}{CS}
\usepackage{braket}
\newcommand{\inv}{\text{inv}}

\begin{document}

%
\begin{titlepage}
\begin{center}
~\\[2cm]
{\fontsize{28pt}{0pt} \mytitlefont Anomaly Matching Across Dimensions and Supersymmetric Cardy Formulae}

~\\[1.25cm]

Kantaro Ohmori\,$^{1,2}$ and\, Luigi Tizzano\,$^{2,3}$\hskip1pt
~\\[0.5cm]
{$^1$~{\it Department of Physics, Faculty of Science,
University of Tokyo, Bunkyo, Tokyo 113-0033, Japan}}
~\\[0.15cm]
{$^2$~{\it Simons Center for Geometry and Physics, SUNY, Stony Brook, NY 11794, USA}}
~\\[0.15cm]
{$^3$~{\it Physique Th\'eorique et Math\'ematique and International Solvay Institutes
Universit\'e Libre de Bruxelles; C.P. 231, 1050 Brussels, Belgium}}
~\\[1.25cm]
\end{center}
\noindent 
't Hooft anomalies are known to induce specific contributions to the effective action at finite temperature. We present a general method to directly calculate such contributions from the anomaly polynomial of a given theory, including a term which involves a $U(1)$ connection for the thermal circle isometry. Based on this observation, we show that the asymptotic behavior of the superconformal index of $4d$ $\mathcal{N}=1$ theories on the ``second sheet" can be calculated by integrating the anomaly polynomial on a particular background. The integration is then performed by an equivariant method to reproduce known results. Our method only depends on the anomaly polynomial and therefore the result is applicable to theories without known Lagrangian formulation. We also present a new formula that relates the behavior of $6d$ $\mathcal{N}=(1,0)$ SCFTs on the second sheet to the anomaly polynomial.

\vfill 
\begin{flushleft}
December 2021
\end{flushleft}
\end{titlepage}
%
		
	
\setcounter{tocdepth}{3}
\renewcommand{\cfttoctitlefont}{\large\bfseries}
\renewcommand{\cftsecaftersnum}{.}
\renewcommand{\cftsubsecaftersnum}{.}
\renewcommand{\cftsubsubsecaftersnum}{.}
\renewcommand{\cftdotsep}{6}
\renewcommand\contentsname{\centerline{Contents}}
	
\tableofcontents


\section{Introduction}
Global symmetries and their 't Hooft anomalies play an essential role in the study of quantum field theory. These can be leveraged to understand various dynamical properties of RG flows in the form of non-renormalization theorems and anomaly matching conditions \cite{Adler:1969er, tHooft:1979rat}.

Anomalies are traditionally studied in the vacuum state. Yet in recent years there has been a surge of interest in exploring what kind of constraints they impose on thermal states \cite{Fukushima:2008xe, Son:2009tf, Neiman:2010zi,Banerjee:2012iz, Jensen:2012jy, Golkar:2012kb, Jensen:2012kj, Jensen:2013kka}, leading to the discovery of new non-dissipative transport phenomena.

A standard way to describe 't Hooft anomalies is to couple a given theory to background gauge fields $\cA$ for the global symmetries or to a background metric $g_{\mu\nu}$. Anomalies arise whenever it is \emph{not} possible to introduce local counterterms for the background fields in such a way that the effective action $\cW_\QFT[\cA]=-\log Z[\cA]$ is invariant under background gauge transformations or coordinate reparametrizations.\footnote{We use $Z[\cA]$ to denote the partition function as a function of all background fields $\cA$ obtained from the functional integral of the Euclidean action over all dynamical fields. Our formalism also applies to theories without a Lagrangian description as long as $Z[\cA]$ exists.} 

Here we will be interested in placing an even $d$-dimensional quantum field theory on a thermal background i.e. on a manifold which is topologically the direct product of a thermal circle and a base (space) manifold. Assuming a thermal gap, we can reduce the theory over the circle and view $\cW_\QFT[\cA]$ as a local generating functional on the $d-1$-dimensional base manifold. Whenever a theory has a 't Hooft anomaly, the anomalous contribution to $\cW_\QFT[\cA]$ give rise to Chern--Simons contact terms on the base manifold whose coefficients are fixed uniquely in terms of the underlying 't Hooft anomaly coefficients. This phenomenon is sometimes referred to as anomaly matching across dimensions. 

Additionally, there are also Chern--Simons terms originating from \emph{gauge invariant} contributions to the effective action in $d$ dimensions. Their coefficients were calculated in several examples \cite{Landsteiner:2011cp,Landsteiner:2011iq,Loganayagam:2012pz} where it was found that they are still proportional to 't Hooft anomalies coefficients appearing in the anomaly polynomial of the $d$-dimensional theory. Another puzzling feature of these terms is that they appear in $\cW_\QFT[\cA]$ at two orders of derivatives lower than expected. It was proposed in \cite{Jensen:2012kj, Jensen:2013rga} that, assuming smoothness of the partition function on a background with a conical singularity, the coefficients of this class of terms can be determined non-perturbatively. However, quantum field theory on a singular spacetime is quite subtle because of the expected localized states near the singularity whose decoupling effects are not well understood.

In this paper we present a new non-perturbative way to fix the coefficients of the aforementioned Chern--Simons contact terms which does not involve singular geometries. Our main point is that, when studying an anomalous theory on a thermal background, we should take into account that the metric has a thermal circle that is non-trivially fibered over the base manifold. This in turn crucially modifies the computation of the anomaly polynomial and gives a new explanation for some formulas appearing in \cite{Jensen:2013rga}. We also show that there is a systematic procedure to obtain all the Chern--Simons contact terms in $d-1$ dimensions directly from the anomaly polynomial.

A particularly interesting application of anomalies in a thermal state is in the context of supersymmetric quantum field theories. Any $4d$ $\cN=1$ theory with a continuous R-symmetry can be placed on a background with topology $\cM_3 \times S^1$, where $\cM_3$ is a Seifert 3-manifold, while preserving at least one time independent supercharge \cite{Dumitrescu:2012ha, Klare:2012gn}. An $S^3\times S^1$ supersymmetric background preserves $4$ supercharges and allows us to define a special protected observable which is a refined Witten index (as in \cite{Witten:1982df}) counting supersymmetric states in the Hilbert space on $S^3$. For theories with $\cN=1$ superconformal invariance, this is known as the superconformal index \cite{Romelsberger:2005eg, Kinney:2005ej}. (See \cite{Rastelli:2016tbz,Gadde:2020yah} for some modern reviews.) It was established in \cite{DiPietro:2014bca} that the $\cN=1$ superconformal index at ``high-temperature" is described by an effective field theory whose local functional in three dimensions consists of a supersymmetrized version of the Chern--Simons contact terms discussed above. The high-temperature asymptotics is governed by a linear combination of the conformal 't Hooft anomaly coefficients $a$ and $c$. Interestingly, a similar behavior was discovered long time ago in the high-temperature limit of 2d (non-supersymmetric) CFTs in \cite{Cardy:1986ie}.

 The $\cN=1$ superconformal index is a \emph{holomorphic} function of the angular momentum fugacities which we denote here by $\omega_1$ and $\omega_2$. Following the analysis of \cite{Closset:2013vra} we can identify $\omega_1$ and $\omega_2$ with the complex structure moduli of a space with topology $S^3\times S^1$. However, the index is not periodic if we shift $\omega_1$ or $\omega_2$ by $2\pi i$ and instead takes values in a multiple cover of the space of complex structure deformations of $S^3\times S^1$ \cite{Copetti:2020dil}.

This property allows us to explore different ``sheets" of validity of the refined index \cite{Choi:2018hmj, Cabo-Bizet:2018ehj, Kim:2019yrz, Cabo-Bizet:2019osg}, by shifting $\omega_1$ or $\omega_2$ by $2\pi i n_0$ where $n_0$ is an integer labeling different complex sheets. The possible values of $n_0$ depend on the R-charge assignment of fundamental fields in the supersymmetric theory of interest. 
For $\cN=1$ theories, $n_0=0$ denotes the first sheet where the original analysis of \cite{DiPietro:2014bca} applies. On the second sheet, where $n_0=1$, the Cardy limit leads to a completely different asymptotics which is governed by a new effective field theory.
A detailed analysis of the $3d$ supersymmetric Chern--Simons contact terms on the second sheet was carried out in \cite{Cassani:2021fyv}.\footnote{Note that there is also a ``complex conjugate" sheet for $n_0=-1$ which is described by the same effective field theory.} Complementary studies of the superconformal index have led to similar conclusions about the asymptotic behavior in \cite{Amariti:2021ubd,ArabiArdehali:2021nsx}. 

Under certain assumptions that we discuss in the main text, the authors of \cite{Cassani:2021fyv} have obtained a universal formula that contains all the analytic terms contributing to the Cardy limit of $\cN=1$ theories on the second sheet. One of the main point that we will address is that the same formula can be derived by evaluating the anomaly polynomial of the $\cN=1$ theory in a thermal background. This amounts to performing an explicit integration of the form
\begin{equation}
    \log \mathcal{I} = \int_{\cY_6}2\pi i P^{(6)}[\mathcal{B}]+\mathcal{O}(\beta^0)\ec
\end{equation}
where $P^{(6)}[\mathcal{B}]$ is the anomaly polynomial 6-form and $\beta \equiv \frac{1}{T}$ is the inverse temperature. The particular background $\mathcal{B}$, the geometry $\cY_6$ and the specific boundary conditions that we impose will be discussed in the later sections.
Our approach is non-perturbative; it does not rely on the effective field theory description and therefore it is also valid for strongly coupled $\cN=1$ theories.

In our study of supersymmetric theories on the second sheet we will integrate the anomaly polynomial equivariantly, following \cite{Bobev:2015kza}. It should be noted that a similar idea was applied to this problem already in \cite{Nahmgoong:2019hko}. Our analysis justifies this approach and explains why it is related to the reduction of anomalies in a thermal state.

Finally, we apply our ideas to six-dimensional $\cN=(1,0)$ SCFTs and present an asymptotic formula for the index on the second sheet even though the relevant $S^5\times S^1$ supersymmetric background and the structure of $5d$ supersymmetric counterterms have not been worked out yet. It would be nice to obtain a complete understanding of this problem and we hope to report about it in a future work.

This paper is organized as follows: in section \ref{terEFT} we discuss various aspects of anomalies in thermal field theory, introduce three-dimensional Chern--Simons contact terms and propose our main non-perturbative formula to fix their coefficients. In section \ref{CardySCFT} we focus on the $\cN=1$ superconformal index. We review the notion of ``sheets" and their geometric interpretation. Lastly, we apply our ideas to compute all the analytic terms contributing to the Cardy limit of $\cN=1$ theories on the second sheet. In section \ref{6dSCFTs} we consider six dimensional $\cN=(1,0)$ SCFTs and use our main formula to compute their Cardy limit on the second sheet. Some mathematical details about equivariant integration are collected in appendix \ref{equivint}.

\section{Thermal Effective Field Theory and Anomalies}\label{terEFT}
Finite temperature effects in quantum field theory are most conveniently studied by placing the theory on a Euclidean spacetime with topology $\cM_{d-1} \times S^1$ where $S^1$ represents a thermal circle of radius $\beta \equiv {\frac1T}$. In this work we will focus on even spacetime dimension $d$ theories that, at sufficiently high temperature, have no gapless degrees of freedom on $\bR^{d-1}$. 

In order to compute equilibrium correlation functions we can further activate background fields for various global symmetries or the metric field. In the high-temperature $\beta \to 0$ limit we can study equilibrium observable by dimensionally reducing the theory along the thermal circle and obtain a local analytic functional of the background fields on $\cM_{d-1}$ which we denote by $\cW_{\QFT}$. Note that this holds on the assumption on the existence of a gap in the reduced theory.\footnote{Note that in certain supersymmetric examples the dimensional reduction on $S^1$ is gapless on $\bR^3$ because of extended supersymmetry. However, the high-temperature holonomy effective potential has isolated minima and hence there are no gapless degrees of freedom in the EFT. 
Following \cite{DiPietro:2016ond}, for Lagrangian QFTs one can always introduce light dynamical fields (lighter than the KK scale $\propto 1/\beta$) in the effective action to define “gapped vacua” as isolated minima of the effective action for the dynamical fields.} 

The functional $\cW_{\QFT}$ admits a derivative expansion in inverse powers of $\beta$ which can be studied systematically. Let us now introduce a four-dimensional metric for $\cM_{3}\times S^1$ of the form
\be\label{KKmetric}
ds^2 = e^{2\Phi}(d\tau + a_i dx^i)^2 + ds_{\cM_3}^2\ec
\ee
where $x_i$, $i=1, 2, 3$ are local coordinates on $\cM_{3}$ and $\tau \simeq \tau + \beta$. The one-form $a$ is often referred to as KK photon. To simplify the discussion we set $\Phi=0$ in what follows. We further assume a $U(1)_{\TT}$ symmetry whose background gauge field $\cA_\TT$ is expanded in components\footnote{Throughout this paper, the subscript $\TT$ stands for ``thermal".} as 
\be\label{KKback}
\cA_\TT = \cA_\tau (d\tau + a_i dx^i) + \cA_i dx^i \ec
\ee
where $\cA_\tau$ is taken along the circle while the $\cA_i$'s take values only along $\cM_3$. At first order in derivatives, the thermal partition function expansion is given by \cite{Banerjee:2012iz}
\be\label{1der}
\begin{split}
\cW_{\QFT}^{[1]}[\cA_\TT] &= -i{C\over 2(2\pi)^2}\beta\int_{\cM_3} \left(\cA_\tau\cA \wedge d\cA + \cA^2_\tau\cA \wedge da + {1\over 3}\cA^3_{\tau}a\wedge da \right)\\ &+ {i k_1\over 2\beta}\int_{\cM_3}\cA \wedge da\ed
\end{split}
\ee
We would like to make several comments about these Chern--Simons terms:
\begin{itemize}
\item The chemical potential $\cA_\tau$ is an arbitrary function on the base manifold $\cM_3$. For this reason the Chern--Simons functionals with overall coefficient $C$ are field dependent and violate invariance under small gauge transformations. However, the four-dimensional theory on $\cM_3 \times S^1$ can have 't Hooft anomalies for various global symmetries and one might expect that, on this background, the non-invariance can be related to such anomalies.
It was indeed shown in \cite{Jensen:2013kka, DiPietro:2014bca} that:
\be\label{k2k3}
C=k_{U(1)^3}\ec
\ee
where $k_{U(1)^3}=\sum_i q^3_i$ is the standard $U(1)^3$ anomaly coefficient for fermions of $U(1)$ charges $q_i$. This is an example of anomaly matching across dimensions. In thermal field theory at high-temperature some observables are determined in terms of the anomaly coefficient of the four-dimensional theory.\footnote{A famous example of such phenomenon is the so-called chiral magnetic effect of \cite{Fukushima:2008xe, Son:2009tf}.} An alternative derivation of \eqref{k2k3} will be discussed in section \ref{aninf}.
\item The most interesting and puzzling contribution to $\cW_{\QFT}^{[1]}[\cA_\TT]$ is the Chern--Simons term with coefficient $k_1$. A priori there is no intuitive reason to expect that $k_1$ is fixed in terms of anomaly matching argument because it multiplies a term which is completely invariant under small gauge transformations. The coefficient $k_1$ has first been computed in free theories \cite{Landsteiner:2011cp} and via holography \cite{Landsteiner:2011iq}, in all these examples it was shown that
\be\label{k1}
k_1 = -{1\over 12} k_{U(1)}\ec
\ee 
where $k_{U(1)}=\sum_i q_i$ is the coefficient of a mixed $U(1)$-gravitational anomaly for fermions of $U(1)$ charges $q_i$. For Lagrangian field theories, evidence for \eqref{k1} has been first shown using perturbation theory in \cite{Golkar:2012kb}. A proof of \eqref{k1} was later established in \cite{DiPietro:2014bca}. See also \cite{Golkar:2015oxw} for an alternative proof. The basic idea is that the coefficient $k_1$ in \eqref{1der} cannot depend on continuous coupling constants. If it were otherwise, we could promote it to a background field and violate invariance under small gauge transformations with no possible anomaly to account for that. In any Lagrangian field theory one can tune the coupling to the free field theory point and compute the coefficient $k_1$ by summing over the KK tower generated by free fields on the thermal circle. By 't Hooft anomaly matching arguments \eqref{k1} is thus established for any value of the coupling.

\item Obtaining a complete non-perturbative proof of \eqref{k1} is more challenging. In \cite{Jensen:2012kj, Jensen:2013rga} it was proposed that $k_1$ can be determined without relying on a Lagrangian description by assuming smoothness of the path integral on a conical geometry. Despite being fixed in terms of a mixed $U(1)$-gravitational anomaly, $k_1$ contributes to the thermal effective action \eqref{1der} at first order in derivatives, i.e. at two orders  lower than expected. The authors of \cite{Jensen:2012kj, Jensen:2013rga} have referred to this phenomenon as ``derivative jump" and motivated the study of thermal field theory on a conical geometry to explain the origin of such unusual behavior.  In these kind of backgrounds it is however not clear how to deal with localized states near the singularity and their decoupling effects.
\end{itemize}
 
In the following we will present a novel way to fix the values of $k_1$ and $C$ using anomaly matching arguments in the background \eqref{KKmetric}. For $k_1$, our argument gives a new non-perturbative proof of the relation \eqref{k1} which bypasses the problem of studying singular geometries and explain some elements of the construction in \cite{Jensen:2012kj, Jensen:2013rga}.

\subsection{Anomaly polynomial, Chern--Simons functional and invertible QFT}\label{aninf}
Let us begin by recalling some elements of perturbative 't Hooft anomalies for continuous global and spacetime symmetries. These are commonly described in terms of a gauge invariant $(d+2)$-form anomaly polynomial $P^{(d+2)}[\cB]$ constructed out of various characteristic classes. Here $\cB$ denotes some background gauge fields for the global symmetries extended to $d+2$ dimensions. A relevant example for us will be that of a four-dimensional theory with a $U(1)$ global symmetry anomaly whose most general anomaly polynomial can be expressed as
\be\label{anpol}
P^{(6)}[\cB] = {k_{U(1)^3}\over 6}c^3_1(\cF)  + {k_{U(1)}\over 24}p_1 c_1(\cF) \ec
\ee
with coefficients $k_{U(1)^3}$ and $k_{U(1)}$ defined in the previous section and $\cB$ denoting a background gauge field for the $U(1)$ global symmetry whose associated field strength is $\cF = d\cB$. The first Chern class $c_1(\cF)$ and first Pontryagin class $p_1$ are defined in the following way
\be\label{classes}
c_1(\cF) = {\cF \over 2\pi}\ec \qquad p_1 = -{1\over 2 (2\pi)^2} \tr (\cR \wedge \cR)\ec
\ee
where $\cR$ is the Riemann curvature 2-form and the trace is performed over $SO(4)$ frame indices.
Given the anomaly polynomial \eqref{anpol}, we associate it to a Chern--Simons action functional on a closed 5-manifold $\cN_5$ with background $\hat\cB$
as:\footnote{In the rest of this section, we use symbols with hats to mean once-extended backgrounds, i.e.\ backgrounds on $\cN_5$, or $\cN_4$ when $\cN_5 = \cN_4\times S^1$, while we use symbols with tildes to mean twice-extended backgrounds, i.e.\ backgrounds on $\cY_6$, or $\cY_5$ when $\cY_6 = \cY_5 \times S^1$. }
\be\label{SCS}
S_{\CS}[\cN_5,\hat\cB] = 2\pi \mathrm{i} \int_{\cY_6} P^{(6)}[\tilde{\cB}]\ec
\ee
where $\cY_6$ is a 6-manifold with boundary $\partial \cY_6 = \cN_5$, and $\tilde{\cB}$ is an extension of $\hat\cB$ onto $\cY_6$. 
A subtle, yet important in this paper, point is that a neighborhood of $\partial \cY_6 = \cN_5$ should be isometric to $\cN_5 \times [0,1)$, called the collar region. Also, we demand that the extended background $\tilde{\cB}$ in the region does not contain the $\mathrm{d}x_6$ component and is independent of $x_6$, where $x_6$ is the coordinate on the $[0,1)$ factor. Given this, if we have two different extensions $(\cY_6,\tilde{\cB})$ and $(\cY_6',\tilde{\cB}')$ of $(\cN_5,\hat{\cB})$, the difference of Chern--Simons functionals is
\begin{equation}\label{CSdiff}
    2\pi \mathrm{i} \left(\int_{\cY_6} P^{(6)}[\tilde{\cB}] - \int_{\cY_6'} P^{(6)}[\tilde{\cB}']\right)
    = 
    2\pi \mathrm{i} \int_{\cY_6\cup_{\cN_5} \bar{\cY_6}'} P^{(6)}[\tilde{\cB}'']\ec
\end{equation}
where $\cup_{\cN_5}$ denotes gluing along the boundary $\cN_5$ and $\tilde{\cB}''$ is the background on $\cY_6 \cup_{\cN_5} \bar{\cY_6}'$ obtained by gluing $\tilde\cB$ on $\cY_6$ and $\tilde\cB'$ on $\cY_6'$.
Note that because of the collar region constraint, both the glued metric and the glued background $\hat\cB''$ are smooth on $\cY_6\cup \bar{\cY_6}'$.
This smoothness enables us to use Chern--Weil theory and the index theorem built on top of it, concluding that the ambiguity \eqref{CSdiff} is an integer multiple of $2\pi \mathrm{i}$, ensuring that $e^{-S_{\CS}[\cN_5,\cB]}$ is well-defined.

A modern viewpoint on the anomaly inflow is to regard the action functional \eqref{SCS} as an \emph{invertible QFT}.\footnote{The invertible QFT corresponding to a perturbative anomaly is not topological, as it depends on the non-topological data of metric and backgrounds.}
An invertible QFT is a Eucledian QFT whose Hilbert space $\cH_{\cM_4}^\inv$ for any closed 4-manifold $\cM_4$ is one-dimensional. When two 4-manifolds $\cM_4$ and $\overline{\cM}_4'$ are bounded by a 5-manifold $\cN_5$, the invertible QFT defines a evolution map along $\cN_5$:
\begin{equation}
    U(\cN_5):\cH_{\cM_4}^\inv \to \cH_{\cM_4'}^\inv\ed
\end{equation}
In particular, if $\cN_5$ is closed,  $U(\cN_5)$ defines a number multiplying on $\cH_{\varnothing}^\inv$. Formally, we pick a unit basis $\ket{0} \in \cH_{\varnothing}^\inv$ and the partition function is 
\begin{equation}\label{invZ}
    Z^\inv[\cN_5] = \braket{0|U(\cN_5)|0}\ed
\end{equation}
When we take a global symmetry into consideration, the invertible QFT also defines a Hilbert space $\cH_{\cM_4,\cB}^\inv$ for a pair of $4$-manifold and the gauge equivalence class of the background $\cB$ on $\cM_4$. It also includes the evolution map $U(\cN_5,\cB)$ for a given pair $(\cN_5,\cB)$.
Then, the partition function \eqref{invZ} should reproduce the Chern--Simons action functional \eqref{SCS}:
\begin{equation}
    Z^\inv[\cN_5,\cB] = e^{iS_{\CS}[\cN_5,\cB]}\ed
\end{equation}

The anomaly inflow picture in this framework is that, the anomalous $4d$ theory $\cT$ should be regarded as a boundary condition of the invertible QFT. 
We let $\ket{\cT[\cM_4]}\in \cH_{\cM_4}^\inv$ denotes the corresponding boundary state.
Then, the gauge invariant partition function of the anomalous 4d theory is non-canonically defined by picking a state $\ket{R[\cM_4,\cB]}\in\cH_{\cM_4,\cB}^\inv$ as:
\begin{equation}\label{ZT}
    Z^{\cT}[\cM_4,\cB] = \braket{\cT[\cM_4,\cB]|R[\cM_4,\cB]}.
\end{equation}
A common choice of the reference state $\ket{R[\cM_4,\cB]}$ is
\begin{equation}
    \ket{R[\cM_4,\cB]} = \ket{\psi[\cN_5],\hat{\cB}} := U[\cN_5,\hat{\cB}] \ket{0},
\end{equation}
where $\cN_5$ is a manifold bounding $\cM_4$ and $\hat{\cB}$ is an extension of the background.\footnote{This choice is available only when $(\cM_4,\cB)$ is null-bordant, and \eqref{ZT} is more general. This further ambiguity is because of the possibility of adding counter term on the boundary $\cM_4$ composed of the bordism invariants.} Note that the partition function \eqref{ZT} for the $5d/4d$ combined system is gauge invariant.

\subsection{Circle Reduction of the Invertible QFT}\label{redinvertible}
Since we are interested in a thermal background, we study the theory on a spacetime manifold which is locally $\cM_4= \cM_3 \times S^1$.
It is natural to demand that the $5d$ bulk also has a thermal circle. Reducing a bulk $5d$ invertible QFT
on the thermal circle, gives rise to a $4d$ invertible QFT.
We would now like to determine such $4d$ invertible QFT. 

For this purpose, we start with a compact $5d$ manifold $\cN_5 = \cN_4 \times S^1$. 
After determining the $4d$ invertible QFT, we can let the $4d$ manifold $\cN_4$ have a boundary.
From the definition \eqref{SCS}, we pick $\cY_6$ to be topologically $\cY_6 = \cN_4 \times D^2$. 
As emphasized below \eqref{SCS}, \emph{the $D^2$ factor must be given a cigar metric}, and not a flat disk metric, in order to give $\cY_6$ a collar region.
Explicitly, we take the metric on $\cY_6$ to be
\begin{equation}\label{L6metric}
    ds_{\cY_6}^2 = dr^2 + g(r)^2 (d\tau + \hat a_I dx^I)^2 + ds^2_{\cN_4}\ec
\end{equation}
where $0\le r\le 1$ is the radial coordinate of $D^2$, $r=1$ is the boundary, and that the function $g(r)$ should satisfy $g(r) =1 $ in the collar region and $g(r=0) = 0$ so that the metric is smooth at the tip $r=0$. The index $I$ runs for the coordinates on $\cN_4$.
To determine the $4d$ invertible QFT, we need integrate over $D^2$.

We now briefly discuss our notations and a few  preparatory calculations.
First, we introduce a chemical potential for the $U(1)$ background field, i.e.\ a non-trivial component $\cA_\tau$ along the imaginary time direction. To be explicit, we take the background gauge field $\hat\cA_\TT$\footnote{Throughout this paper, the subscript $\TT$ stands for ``thermal".}
on $\cN_5=\cN_4\times S^1$ to be 
\be\label{hatAT}
\hat{\cA}_\TT = \hat{\cA}_\tau (d\tau + \hat{a}_I dx^I) + \hat{\cA}_I dx^I \ec
\ee
where the hat symbol denote (extended) bulk  fields of the $5d$ invertible QFT.
We would like to further extend this onto $\cY_6 = \cN_4\times D^2$. We thus set the extended background to be
\be\label{tildeAT}
\tilde{\cA}_\TT = f(r) \hat{\cA}_\tau (d\tau + \hat{a}_I dx^I) + \hat{\cA}_I dx^I \ed
\ee
The smooth function $f(r)$ is such that $f(r=0) = 0$ so that $\tilde{\cA}_\TT$ is well-defined at the origin, and also $f(r)=1$ in the collar region.
The curvature of this extended background is
\begin{equation}\label{tildeF}
    \tilde\cF_\TT = d \tilde \cA_\TT = \left(d(f(r)\hat{\cA}_\tau(d\tau+\hat{a})) + d \hat\cA\right)\ed
\end{equation}

We are now ready to perform the fiber integration of the $U(1)^3$ 't Hooft anomaly polynomial:
\begin{equation}\label{c1cube_integrate}
    \begin{split}
        2\pi\mathrm{i}\int_{\cN_4\times D^2}\frac{k_{U(1)^3}}{6}c_1^3(\tilde\cF_\TT) &= 2\pi \mathrm{i}\, \frac{k_{U(1)^3}\beta}{6 (2\pi)^3} \int_{\cN_4} \left(3\hat\cA_\tau(d\hat\cA)^2 + 3\hat\cA_\tau^2 d\hat\cA da +  \hat\cA_\tau^3 (da)^3\right).
    \end{split}
\end{equation}
We have expended $\tilde{\cA_\TT}^3$ and used the boundary conditions for $f(r)$ explained below \eqref{tildeAT} to evaluate  $\int dr f'(r) f(r)^n =  \int df(r) f(r)^n = \frac{1}{n}$.\footnote{If one naively integrates the local expression $S_{\CS} \sim \hat\cA_\TT  \hat\cF_\TT^2$ over $S^1$ assuming that $\hat{\cA}_\tau$ is constant, one would not get the factor $3$ in the $\cA_\tau (d\cA)^2$ term. However such a calculation is valid only when $\hat\cA_\TT$ is a globally well-defined 1-form and in such cases $S_{\CS} =0$.} Next, we calculate the fiber integration of $p_1c_1(\tilde\cF_\TT)$. Note that $\cY_6$, which is topologically a  product $\cN_4\times D_2$, has tangent bundle $T\cY_6$ splitting into the direct sum $T\cN_4\oplus TD_2$. Whenever we compute the first Pontryagin class $p_1$, we need to consider the splitting pattern (Whitney's splitting formula):
\begin{equation}\label{whitney}
    p_1(T(\cN_4\times D^2)) = p_1 (T\cN_4 \oplus TD^2) = p_1(T\cN_4) + p_1(TD^2)\ed
\end{equation}
In addition, the Pontryagin class on 2-manifold is its Euler class squared: $p_1(TD^2) = \chi(TD^2)^2$,
where the Euler class is explicitly\footnote{To check the normalization, we can integrate $\chi$ over  $D^2$ and obtain $1$. This is the correct Euler number of the Disk. The Euler number of an open surface receives a contribution from the boundary extrinsic curvature. However, with a cigar metric the extrinsic curvature on the boundary vanishes.}
\begin{equation}\label{chibulk}
\chi(TD^2) = {1\over 4\pi} \cR(TD^2) = \frac{d(g(r)(d\tau+\hat a))}{\beta}.
\end{equation}
The first term is just the usual formula for the Euler class of a surface, and the second term comes from the twisting of the $D^2$ fiber caused by $g(r) a$.

We can now perform the fiber integration of $p_1c_1(\tilde\cF_\TT)$.
To simplify the calculation, we pick the two interpolation functions $f(r)$ and $g(r)$ such that where $1/2 \le r \le 1$, $f(r)$ interpolates between $0$ and $1$ while $g(r)=1$, and where $0\le r\le 1/2$ $g(r)$ interpolates between $0$ and $1$ while $f(r) = 0$. In particular $f(r) dg(r) = 0$.
Then, the fiber integration goes as follows:
\begin{equation}\label{p1c1_integrate}
    \begin{split}
        2\pi\mathrm{i}\int_{\cN_4\times D^2}&\frac{k_{U(1)}}{24}p_1(T\cY_6) c_1(\tilde\cF_\TT) = 
        2\pi \mathrm{i}\,\frac{k_{U(1)}}{24(2\pi)}\int_{\cN_4\times D^2}(p_1(T\cN_4)+\chi(D^2)^2)\tilde{\cF_\TT}\\
        &= 2\pi \mathrm{i}\,\frac{k_{U(1)}}{24(2\pi)}\int_{\cN_4\times D^2}\left(2\frac1{\beta^2}dg(r)d\tau\, d(g(r)\,a\,)+\frac{1}{\beta^2}(d(g(r)\, a))^2\right) \tilde{\cF_\TT}\\
        &= 2\pi \mathrm{i}\, \frac{k_{U(1)}}{24 (2\pi)} \int_{\cN_4} \left(\frac1{ \beta}d\hat a \,  d\hat\cA + \beta \hat{A}_\tau p_1(T\cN_4)  \right).
    \end{split}
\end{equation}
In summary, we have obtained the formula
\begin{equation}
    S_{CS}[\cN_4\times S^1, \hat\cA_\tau, \hat\cA, \hat a] = 2\pi\mathrm{i}\,\int_{\cN_4}Q_{\mathsf{T}}^{(4)},
\end{equation}
where the integrand $Q_\TT^{(4)}$ is
\begin{equation}\label{Q4}
\begin{split}
    Q_{\TT}^{(4)} &= 
         \frac{k_{U(1)^3}}{2 (2\pi)^3} \beta \left( \hat\cA_\tau (d\hat \cA)^2 + \hat\cA_\tau^2 d\hat\cA d\hat a + \frac13 \hat\cA_\tau^3 (d\hat a)^2 \right) \\
         &\hspace{5em}  + \frac{k_{U(1)}}{24 (2\pi)} \left(\frac1{\beta}d\hat a \, d\hat\cA + \beta \hat{\cA}_\tau p_1(T\cN_4)\right),
\end{split}
\end{equation}
with
\begin{equation}\label{hatFTT}
    \hat\cF_\TT = 
     \left(d(\hat{\cA}_\tau\hat{a}) + d \hat\cA\right).
\end{equation}
The functional $Q_\TT^{(4)}$ is invariant under background gauge transformations
\begin{equation}\label{Aa_gauge}
    \hat\cA \to \hat\cA + d \hat{\lambda}\ec\quad \hat a \to \hat a + d \hat\lambda_a\ec
\end{equation}
and coordinate transformations on $\cN_4$.
$Q_\TT^{(4)}$ can be regarded as a Lagrangian defining a $4d$ invertible QFT coupled to the periodic scalar background $\hat{A}_\tau$ and one-form backgrounds $\hat \cA$ and $\hat a$. 
Invertible QFTs involving periodic scalar backgrounds are discussed e.g.\ in \cite{Cordova:2019jnf,Cordova:2019uob}.
The periodicity $\beta \hat{A}_\tau \cong \beta\hat{A}_\tau + 2\pi$ requires a quantization conditions on $k_{U(1)^3}$ and $k_{U(1)}$ in $Q_{\TT}^{(4)}$, which is here guaranteed by the well-definedness of the $5d$ Chern--Simons functional $S_{\CS}$.

\subsection{The \texorpdfstring{$3d$}{3d} Effective Action}\label{effact}
Now we consider the boundary theory of the invertible QFT.
On $\cM_4 = \cM_3 \times S^1$, we have a chiral 4d theory defining the boundary state $\ket{\cT[M_4]}$ of the $5d$ invertible QFT.
Upon, circle reduction, we obtain the boundary state $\ket{\cT[M_3]}_{Q_\TT^{(4)}}$ of the $Q_\TT^{(4)}$ invertible QFT in $4d$.
This invertible QFT does admit a symmetry-preserving (Neumanm) gapped boundary $\ket{N[\cM_3]}_{Q_\TT^{(4)}}$\footnote{The scalar background $\cA_{\tau}$ has to be single-valued on the boundary, in order for the boundary to have a gap everywhere. In other words the $4d$ invertible QFT represents the anomaly for the "-1"-form symmetry of the periodic identification of the scalar background\cite{Cordova:2019jnf,Cordova:2019uob}.}
and our assumption about the thermal gap means that $\ket{\cT[M_3]} = e^{-S_\text{3d}[\cM_3]}\ket{N[M_3]}_{Q_\TT^{(4)}}$ for some well-defined local $3d$ classical action $S_\text{3d}[\cM_3]$, which is not universal. 
Then, the universal contribution is given by the $Q_\TT^{(4)}$ invertible theory inner product
\begin{equation}\label{Wuniv}
    e^{-\cW_\text{QFT}^{[1]}} = \braket{N|R}_{Q_\TT^{(4)}}\ec
\end{equation}
where the reference state $\ket{R}_{Q_{\TT}^{(4)}}$ is the circle reduction of the state $\ket{R}$ in the $5d$ invertible QFT.
If we choose to extend the manifold $\cM_3$ to $\cN_4$ ($\partial \cN_4 = \cM_3$), i.e.\ $\ket{R} = U(\cN_5)\ket{0}$, we get the universal part of the effective action:
\begin{equation}\label{WunivN4}
    \cW_\text{QFT}^{[1]} = 2\pi \mathrm{i} \int_{\cN_4\times D^2}P^{(6)}[\hat{\cA}_\tau,\hat\cA,\hat a] = 2\pi \mathrm{i} \int_{\cN_4}Q_\TT^{(4)}[\hat{\cA}_\tau,\hat\cA,\hat a]\ed
\end{equation}

To make contact with the expression \eqref{1der}, we set the chemical potential $\hat \cA_\tau$ to be constant over $\cN_4$ and do integration by parts:
\be\label{1der_2}
\begin{split}
\cW_{\QFT}^{[1]}[\cA_\TT] &= -i{k_{U(1)^3}\over 2(2\pi)^2}\beta\int_{\cM_3} \left(\cA_\tau\cA \wedge d\cA + \cA_\tau \cA \wedge d(\cA_\tau a) + {1\over 3}\cA^3_{\tau}a\wedge da \right)\\ &+ {i k_{U(1)}\over 24\beta}\int_{\cM_3}\cA \wedge da \ec
\end{split}
\ee
which agrees with \eqref{1der}.\footnote{ Here we ignore a gravitational Chern--Simons terms of the form $\beta^2 \cA_\tau \CS_{\text{grav}}$ which enters the effective action at higher order in derivatives.}

Let us summarize the main aspects of our proposal:
\begin{itemize}
\item In the high-temperature limit, the \emph{complete} contribution to the thermal effective action at first order in derivative $\cW^{[1]}_{\QFT}[\cA_\TT]$ can be determined uniquely in terms of anomaly matching.
Given a $4d$ quantum field theory on a thermal background we first need to compute the contribution from all anomalies by introducing a six-dimensional geometry $\cY_6 = D_2 \times \cN_4$ and fibered metric \eqref{L6metric}. In terms of these data we obtain a ``thermal" anomaly polynomial six-form given by
\be 
P_{\mathsf{T}}^{(6)}[\cB] = \frac{k_{U(1)^3}}{6}c_1^3(\tilde\cF_\TT)  +\frac{k_{U(1)}}{24}p_1(T\cY_6) c_1(\tilde\cF_\TT) \ed
\ee 
In the high-temperature $\beta \to 0$ limit we can reduce $P_{\mathsf{T}}^{(6)}$ along the $D_2$ fiber and obtain
\be\label{termalmatch}
Q_{\mathsf T}^{(4)}[\hat\cA_\TT] = \int P_{\mathsf T}^{(6)}[\cB] \biggr|_{D_2}\ec
\ee
in this approach $Q_{\mathsf T}^{(4)}[\hat\cA_\TT]$ is the anomaly 4-form polynomial for the high-temperature thermal effective action on $\cM_3$. The equation on the right hand side of \eqref{termalmatch} should be thought of as a formal way to denote that we are integrating along the fibers of $\cY_6$. We can furthermore make a particular choice of extension for $\hat\cA_\TT$ to reproduce the thermal effective action \eqref{1der} and therefore fix the coefficients $k_1$ and $C$. Our argument sets the value of $k_1$ non-perturbatively.

\item The concept of thermal anomaly polynomial $P_{\mathsf{T}}^{(6)}$ and the related ``substitution rule" have already appeared in \cite{Loganayagam:2012pz, Jensen:2012kj, Jensen:2013rga}. However, in those works, one has to require that the thermal effective action is consistent with the ``Euclidean vacuum" (which necessarily implies considering a conical background geometry) and modify the prescription of anomaly inflow to obtain an anomaly polynomial of the right degree. Here we see that no such technical difficulties are introduced and the substitution rule simply follows from the splitting \eqref{whitney}. 

\item The study of anomaly constraints in thermal quantum field theory has a wide range of applicability. Our arguments can be easily adapted to describe theories with different even spacetime dimension. In the rest of this paper we will be mainly interested in $4d$ supersymmetric theories. As we will explain in section \ref{CardySCFT}, the power of supersymmetry together with anomaly matching arguments allow us to fix a natural choice for the extension $\hat\cA_\TT$ in that particular context.

\item Around equation \eqref{chibulk} we relied on $D^2$ having a cigar metric. This causes a ``derivative jump". If one naively treats the $k_{U(1)}$ dependent part of $S_{\CS}$ as $\hat \cA p_1$, one might not expect a contribution from this term upon circle compactification. However, the correct treatment requires to integrate the anomaly polynomial $c_1p_1$ over the cigar (and not over the flat disk), and around the tip of the cigar $p_1$ takes nontrivial values.

\item Finally, we notice that the $D^2$ fiber integration in \eqref{WunivN4} could be done equivariantly. That is, we could promote both $\tilde\cF_\TT$ and $\chi$ to the their equivariant version, and then use the Atiyah–Bott–Berline–Vergne equivariant localization formula to localize the contribution at the tip of the cigar $D^2$. The collar condition ensures that the boundary does not give rise to additional contribution when applying the formula. See Appendix~\ref{equivint} for some additional details on equivariant integrals. In the next section, we perform the equivariant integration for the whole $D^2\times \cN_4$ integration with a particular background corresponding to the superconformal index.
\end{itemize}

\section{The Cardy Limit of Superconformal Theories}\label{CardySCFT}
Any $4d$ $\cN=1$ superconfomal field theory has a $U(1)_R$ global symmetry and a special protected observable, the superconformal index $\cI$ \cite{Romelsberger:2005eg, Kinney:2005ej}, defined by \be\label{WittenInd}
\cI = \Tr_{\cH_{\textrm{BPS}}} (-1)^F e^{-\omega_1(J_1 + \half R)} e^{-\omega_2(J_2 + \half R)}\ed
\ee
The trace is performed over the Hilbert space $\cH_{\textrm{BPS}}$ of supersymmetric ground states defined in terms of 
a single $\cN=1$ supercharge $\cQ$ and a Hamiltonian proportional to $\{\cQ,\cQ^\dagger\}$. This can be thought of as a Witten index for ${1\over 4}$-BPS local operators refined by two additional fugacities $\omega_1, \omega_2$ for the spins $J_1, J_2$ corresponding to diagonal combinations of the Cartans of $SU(2) \times SU(2)$. Bosonic states have $J_1, J_2 \in \bZ$ while fermionic states have $J_1, J_2 \in \half \bZ$.

In Lagrangian field theories the index \eqref{WittenInd} can be computed exactly using supersymmetric localization of the partition function $Z_{S^3\times S^1}$ on a Euclidean space with topology $S^3 \times S^1$ and periodic boundary conditions for the fermions along $S^1$. From \cite{Closset:2013vra}, we recall that any manifold with topology $S^3 \times S^1$ is characterized by a two-dimensional moduli space of complex structure deformations $\cM_{\textrm{cpx}}$. The supersymmetric partition function $Z_{S^3\times S^1}$ geometrizes the index by identifying its fugacities $\omega_1$ and $\omega_2$ with the complex structure moduli $p$ and $q$, where $p = e^{-\omega_1}$ and $ q = e^{-\omega_2}$. From this point of view, the superconformal index $\cI$ is a \emph{holomorphic} function of the fugacities which naturally accommodates complex values of $\omega_1$ and $\omega_2$.

A crucial observation is that the index \eqref{WittenInd} is \emph{not} a single valued function of $p$ and $q$ but it takes values on a multiple cover of $\cM_{\textrm{cpx}}$. As a result there are different \emph{sheets} of validity, a terminology introduced in \cite{Copetti:2020dil, Cassani:2021fyv} (see also \cite{ArabiArdehali:2019orz}), for $\cI$ which can be accessed by shifting $\omega_1$ and $\omega_2$ by integer multiples of $2\pi i$.

In this paper we will be interested in analyzing the following limit
\be\label{Cardy}
\omega_1,\omega_2 \to 0\ec ~\text{with ${\omega_1 \over \omega_2}$ fixed}\ed
\ee
When $\omega_1$ and $\omega_2$ are real, we can identify
\be\label{realpar}
\omega_1 = {\beta \over \ell} b\ec \quad \omega_2 =  {\beta \over \ell} b^{-1}\ed
\ee
where $\beta$ and $\ell$ are respectively the $S^1$ and the $S^3$ radii and $b$ is a squashing parameter for the metric on $S^3$. It is clear from \eqref{realpar} that, when $\omega_1$ and $\omega_2$ are real, the limit \eqref{Cardy} is the standard high-temperature Cardy limit for conformal field theories \cite{Cardy:1986ie} where ${\beta\over \ell}\to 0$ with squashing parameter $b$ fixed. A new important element appears in the analysis if we consider instead general complex parameters
\be\label{twist}
\omega_1 = {\beta \over \ell} (b + i \sigma_1)\ec \quad \omega_2 =  {\beta \over \ell} (b^{-1}+ i \sigma_2)\ec
\ee
where $\sigma_1$ and $\sigma_2$ parametrize a special $S^3\times S^1$ supersymmetric background characterized by a certain twisting of $S^3$ around $S^1$. We will describe further aspects of this background in section \ref{susybackg}.

To probe the second sheet of the refined index we thus need to shift $\omega_1 \to \omega_1 + 2\pi i$ while keeping $\omega_2$ fixed. It follows from \eqref{WittenInd} that we can rewrite the index as 
\be\label{Rindex}
\cI(\omega_1, \omega_2) = \Tr_{\cH_{\textrm{BPS}}} e^{-i \pi R} e^{-\omega_1(J_1 + \half R)} e^{-\omega_2(J_2 + \half R)}\ed
\ee
A factor of $e^{-i \pi R}$ is now replacing $(-1)^F$ in \eqref{WittenInd}. We also notice that, looking at \eqref{twist}, a shift by integer multiples of $2\pi i$ of any $\omega_i$ corresponds geometrically to a shift in the twisting parameters $\sigma_i$. The transformed index is still only sensitive to supersymmetric ground states but we expect a different pattern of cancelations to occur. Indeed, compared to \eqref{WittenInd}, the above observable has a much faster growth behavior in the limit \eqref{Cardy}. This property has been studied very extensively in \cite{Choi:2018hmj, Honda:2019cio, ArabiArdehali:2019tdm, Kim:2019yrz, Cabo-Bizet:2019osg} to clarify the microstate counting for large AdS$_5$ black holes, a subject that has received renewed interest in view of recent progress \cite{Hosseini:2017mds,Benini:2018ywd,Choi:2018hmj,Cabo-Bizet:2018ehj}.

The study of the high-temperature $\beta \to 0$ limit of \eqref{Rindex} can be approached using a supersymmetric version of the thermal effective field theory described in section \ref{terEFT}. Indeed, this strategy was already adopted to examine the Cardy limit of the index (on the first sheet) in \cite{DiPietro:2014bca}. Recently, the authors of \cite{Cassani:2021fyv} have generalized this idea to capture the high-temperature behavior of the index on the second sheet. Consistency with $3d$ $\cN=2$ supersymmetry requires that the contact terms appearing in the effective action are, one-loop exact, supersymmetric versions of the Chern--Simons terms from section \ref{terEFT}.

Since an exact analytic expression of \eqref{Rindex} is available, one can implement the Cardy limit directly at the level of the unitary matrix integral \cite{GonzalezLezcano:2020yeb, Amariti:2021ubd,ArabiArdehali:2021nsx}. It was furthermore argued in those works that the high-temperature expansion on the second sheet truncates at $\cO(\beta)$ up to exponentially suppressed terms. 

Following \cite{Cassani:2021fyv}, we assume that all the local terms that might contribute to the effective action, beyond the supersymmetric Chern--Simons contact terms, are true D-terms.\footnote{See also \cite{Ardehali:2021irq} for related discussions.} Moreover, any such D-term must evaluate to zero on the supersymmetric background implying
that we can obtain a powerful universal formula\footnote{Note that the superconformal index $\cI$ and the partition function $Z_{S^3 \times S^1}$ differs by a contribution which is known as the Casimir energy \cite{ArabiArdehali:2015iow, Assel:2015nca}
\be\nonumber
\log Z_{S^3 \times S^1} = -\beta E_{C} + \log \cI\ed
\ee
In the small-$\beta$ expansion the Casimir energy $E_C$ is the leading contribution at order $\cO(\beta)$ and should be contained in the formula \eqref{ZCformula}. As we will see in section \ref{proof}, the contribution from $E_C$ is directly visible in our anomaly matching based approach.} 
which reads:

\be\label{ZCformula}
\log \cI = {(\omega_1+\omega_2+2\pi i)^3\over 48 \omega_1\omega_2} k_{RRR} - {(\omega_1+\omega_2+2\pi i)(\omega_1^2+\omega_2^2 -4\pi^2)\over 48 \omega_1\omega_2} k_{R} + \log |G|  + \cO\left(e^{-{\beta \over \ell}}\right)\ec
\ee
where $k_{RRR}$ and $k_{R}$ are the $U(1)_R$ 't Hooft anomalies coefficients. The $\log |G|$ factor in \eqref{ZCformula} is needed to describe theories with a spontaneously broken 1-form symmetry $G$. We will explore the implications of this term in section \ref{proof}.\footnote{A subtlety related to \eqref{ZCformula} is that this formula is not universally correct on the second sheet. It applies when $\arg(\omega_{1,2})$ are such that the formula yields exponential growth. If $\arg(\omega_{1,2})$ lie outside this range, the asymptotic is not universal and \eqref{ZCformula} can break down. (For $\cN = 4$ super Yang--Mills theory this has been studied in \cite{ArabiArdehali:2021nsx}). In the language of \cite{ArabiArdehali:2019orz}, when the fully deconfined saddle corresponding to \eqref{ZCformula} is exponentially suppressed, other (possible partially deconfined) saddles may take over and dictate the asymptotics.}

Motivated by our analysis in section \ref{terEFT} we would like to find an alternative derivation of \eqref{ZCformula} which does not rely on effective field theory arguments. As we will show in section \ref{proof}, it is in fact possible to determine the high-temperature behavior on the second sheet completely in terms of anomalies. Our results establish the expansion \eqref{ZCformula} non-perturbatively and thereby extend the analysis in \cite{Cassani:2021fyv} to superconformal theories without a weakly coupled description.

\subsection{Twisted \texorpdfstring{$S^3\times S^1$}{S3xS1} Supersymmetric Background}\label{susybackg}
Supersymmetric quantum field theories can be studied systematically on a curved Riemannian manifold $\cM$ by treating the metric $g_{\mu\nu}$ as a component of a background (non-dynamical) supergravity multiplet \cite{Festuccia:2011ws}. See \cite{Dumitrescu:2016ltq} for a review. A supersymmetric background on $\cM$ is characterized by a set of bosonic supergravity background fields which are determined by requiring that all independent supergravity gravitino variation vanish. 

The coupling to background supergravity is governed by the supercurrent multiplet, which includes both the supersymmetry current and the energy-momentum tensor. At linearized level, the curved space Lagrangian $\SL$ differs from the original flat-space one $\mathscr{L}$ by a deformation $\Delta \mathscr{L}$ which is obtained by coupling all the operators in the supercurrent multiplet to their background supergravity partners.  

Here we will exclusively focus on four-dimensional $\cN=1$ theories which possess a supercurrent multiplet, known as the $\cR$-multiplet, characterized by the operators
\be
J^{(R)}_{\mu}\ec\quad S_{\alpha\mu}\ec\quad \tilde{S}^{\alphadot}_{\;\:\mu}\ec\quad T_{\mu\nu}\ec\quad \cF_{\mu\nu}\ec
\ee
where $J^{(R)}_{\mu}$ is the $R$-symmetry current, $S_{\alpha\mu}$ and $\tilde{S}^{\alphadot}_{\;\:\mu}$ are the supersymmetry currents\footnote{Note that there are two independent supersymmetry currents operators. This is due to Euclidean signature where left and right-handed $4d$ spinors are not related by complex conjugation.}, $T_{\mu\nu}$ is the energy-momentum tensor and $\cF_{\mu\nu}$ is a closed two-form which gives rise to a string current $\epsilon_{\mu\nu\rho\lambda}\cF^{\rho\lambda}$. The corresponding background supergravity multiplet contains
\be
\cA^{(R)}_{\mu}\ec\quad\Delta g_{\mu\nu}\ec\quad \Psi_{\alpha\mu}\ec\quad \tilde{\Psi}^{\alphadot}_{\;\:\mu}\ec\quad \cB_{\mu\nu}\ec
\ee
where $\cA^{(R)}_{\mu}$ is an Abelian gauge field, $\Delta g_{\mu\nu}$ is related to the curved background metric $g_{\mu\nu}$ by  $g_{\mu\nu}= \delta_{\mu\nu} +\Delta g_{\mu\nu}$, $\Psi_{\alpha\mu}$ and $\tilde{\Psi}^{\alphadot}_{\;\:\mu}$ are gravitinos and $\cB_{\mu\nu}$ is a Kalb--Ramond two-form gauge field. The supersymmetric deformation $\Delta\SL$ is given by
\be
\Delta {\SL} = - \half \Delta g^{\mu\nu} T_{\mu\nu} + \cA^{(R )\mu} J_\mu^{( R )} + {i \over 4} \ep^{\mu\nu\rho\lambda} \cF_{\mu\nu} \cB_{\rho\lambda} + ({\rm fermions})~.
\ee
A supersymmetric background on a four-manifold $\cM_4$ is defined in terms of a set of data $(g_{\mu\nu}, \cA^{( R)}_\mu,\cB_{\mu\nu})$ which preserve a supercharge~$Q$ or~$\t Q$ of~$R$-charge~$\pm 1$, i.e.\ the supergravity variations~$\delta_Q \Psi_{\mu}$ or~$\delta_{\t Q} {\t \Psi}_\mu$ vanish. This in turn implies that there exists spinor parameters $\zeta_\alpha$ and $\tilde{\zeta}^{\alphadot}$ of opposite $R$-charge $\pm 1$ satisfying certain (generalized) Killing spinors equations which also involve the bosonic background supergravity fields $\cA^{(R)}$ and $\cB$ \cite{Dumitrescu:2012ha, Klare:2012gn}.

In order to analyze the superconformal index $\cI$ we are interested in studying a four-manifold with topology $S^3\times S^1$ and metric 
\be\label{4dmetric}
\begin{split}
ds^2 = d\tau^2 +\, \ell^2(b^2 \cos^2\theta &+b^{-2} \sin^2\theta)d\theta^2\\ &+ b^{-2}\cos^2\theta(\ell d\varphi_1 + \sigma_1 d\tau)^2 + b^{2}\sin^2\theta(\ell d\varphi_2 + \sigma_2 d\tau)^2\ec
\end{split}
\ee
where $\tau \simeq \tau + \beta$, $(\theta, \varphi_1, \varphi_2)$ are local coordinates on $S^3$ subject to $\theta \in (0, {\pi\over 2})$, $\varphi_1\simeq \varphi_1 + 2\pi$ and $\varphi_2\simeq \varphi_2 + 2\pi$. The parameter $b$ controls the squashing of $S^3$ while $\sigma_1$ and $\sigma_2$ control the twisting of $S^3$ around $S^1$. The resulting metric on the four-manifold is therefore non-trivially fibered. It is useful to also introduce a complex Killing vector
\be\label{eq:Kvector}
K = \half\Bigl(-i\partial_\tau+ \ell^{-1}(b+i \sigma_1)\partial_{\varphi_1} + \ell^{-1}\left(b^{-1}+i \sigma_2\right)\partial_{\varphi_2} \Bigr)\ed
\ee
As a one-form $K$ can be expressed as
\be\label{Kone}
K = \frac{1}{2} \left( \left(\frac{\sigma_1}{b} \cos ^2\theta + \sigma_2\,b \sin ^2\theta -i\right)\,d\tau +  {\ell\over b} \cos^2\theta\,d\varphi_1 +  {\ell b} \sin^2 \theta\, d\varphi_2\right) \ed
\ee
It was established in \cite{Dumitrescu:2012ha, Klare:2012gn} that a complex four-manifold with metric \eqref{4dmetric} admits a ``new-minimal" supersymmetric background with two supercharges of opposite $R$-charge. Explicit non-vanishing solutions $\zeta$ and $\t \zeta$ to the Killing spinor equations with metric \eqref{4dmetric}, together with a choice of profile for the background gauge fields $\cA^{(R)}$ and $\cB$  have been fully worked out in \cite{Cassani:2021fyv}. Four our purposes, it will be sufficient to only discuss the explicit expression of $\cA^{(R)}$ which is given by
\be\label{backA}
\begin{split}
\cA^{(R)} &= {1\over 2\ell\sqrt{ b^{-2}{\sin ^2\theta }+b^2 \cos ^2\theta }} \left( \left(2i -{\sigma_1\over b} -b\,\sigma_2\right)d\tau   -\ell\left({\sigma_1\over b} d\varphi_1 - b\sigma_2\, d\varphi_2\right)\right)\\  
&+ \half(d\varphi_1 + d\varphi_2) + {3\over 2}\kappa(\theta)K\ec
\end{split}
\ee
where $\kappa(\theta)$ is an arbitrary function of $\theta$ whose role is not important here.\footnote{One could also introduce an overall conformal factor $\Omega^2(\theta)$ in front of the metric \eqref{4dmetric}. Here we simply set $\Omega(\theta) = 1$. Note that both $\kappa(\theta)$ and $\Omega(\theta)$ are not expected to modify in any way the final result as the partition function $Z_{S^3\times S^1}$ should only depend on complex structure deformations \cite{Closset:2013vra}.} The curved space supersymmetry algebra obtained from $\zeta$ and ${\t\zeta}$ is such that
\begin{equation}\label{susyalg}
\{\delta_\zeta, \delta_{\t \zeta}\} = 2i \cL_{K} + 2K^{\mu}\cA^{(R)}_{\mu} R\ec \qquad\delta^2_\zeta = \delta^2_{\t\zeta} =0 \ec
\end{equation}
where $\cL_{K}$ is a Lie derivative along the Killing vector $K$ and $R$ denotes the $U(1)_R$ charge. Since it will be useful in section \ref{proof} below, we now evaluate explicitly the contraction $2K^{\mu}\cA^{(R)}_{\mu}$ appearing in \eqref{susyalg}. From \eqref{Kone} and \eqref{backA}, we obtain
\be\label{contr}
\mu_R \equiv 2K^{\mu}\cA^{(R)}_{\mu} = \frac{1+ b^2+i b (\sigma_1+\sigma_2)}{2 b \ell} = \half(\omega_1 + \omega_2)\ec
\ee
where in the last equality we used the identification \eqref{twist}. As we have seen in section \ref{CardySCFT}, it is possible to explore different complex sheets of the superconformal index (whenever they are available) by performing a shift $\omega_1 \to \omega_1 + 2\pi in_0$ where $n_0$ denotes the various sheets. For example, in $\cN=1$ superconformal theories, $n_0=0$ is the first sheet while $n_0=\pm 1$ are the second and its complex conjugate sheet. After performing the shift, the parameter $\mu_R$ becomes
\be\label{gammaext}
\mu_R = \half (\omega_1 + \omega_2 + 2\pi i n_0)\ed
\ee
We interpret $\mu_R$ as a $U(1)_R$ chemical potential appearing in the curved space Hamiltonian defined by \eqref{susyalg}. It also follows that, since the superconformal index is a holomorphic function of the fugacities $\omega_1$ and $\omega_2$, $\mu_R$ can be naturally extended to any sheet by analyticity. As we will see below, $\mu_R$ is needed to evaluate the contribution from 't Hooft anomalies of $\cN=1$ theories placed on the $S^3\times S^1$ supersymmetric background presented in this section.

\subsection{New Cardy Formula from Anomaly Matching}\label{proof}
In this section we argue that the Cardy limit of $\cN=1$ SCFTs on the second sheet is completely determined by anomaly matching across dimensions. Our main task will be to evaluate the thermal anomaly polynomial following the prescription from section \ref{terEFT}. We first focus on a simple example, a free chiral multiplet, before discussing the generalization to arbitrary $\cN=1$ SCFTs.

\paragraph{Free Chiral Multiplet}
Consider a theory consisting of a free $\cN=1$ chiral multiplet with $U(1)_R$ charge $0<r<2$. To probe its high-temperature behavior on the second sheet, we need to place the theory on a space with metric \eqref{4dmetric} and activate the supersymmetric background described in section \ref{susybackg}. It follows from formula \eqref{ZCformula} that
\be\label{cardychiral}
\begin{split}
\log \cI_{\textrm{Chiral}} &= {(\omega_1+\omega_2 + 2\pi i n_0)^3\over 48 \omega_1\omega_2}(r-1)^3 -  {(\omega^2_1+\omega^2_2 - 4\pi^2)(\omega_1+\omega_2+ 2\pi i n_0)\over 48 \omega_1\omega_2}(r-1) \\&+ \cO\left(e^{-{\beta \over \ell}}\right) \ec
\end{split}
\ee
where $\log |G| = 1$ since there is no one-form symmetry in this case. We would like to derive the above formula from anomaly matching considerations. A $4d$ chiral fermion has an anomaly polynomial which is given by 
\be\label{anfreechiral}
P^{(6)}[\cB_R] = {1\over 6}c^3_1(\cF_R) - {1\over 24}p_1 c_1(\cF_R) \ec
\ee
where $\cB_R$ is a background gauge field for the $U(1)_R$ and $\cF_R$ is its field strength. The first term in the anomaly polynomial accounts for a cubic $U(1)^3_R$ 't Hooft anomaly, while the second contribution comes from a mixed gravitational-$U(1)_R$ anomaly.  One of the main outcomes from sections \ref{redinvertible} and \ref{effact} is that, the anomaly polynomial \eqref{anfreechiral} receives a subtle geometrical contribution due to the fact that we are studying a theory on thermal background which is a twisted $S^3\times S^1$ fibration. Consequently, we should consider the thermal anomaly polynomial six-form described in section \ref{effact}
\be\label{frechiralT}
P_{\mathsf{T}}^{(6)}[\cB_R] = \frac{1}{6}c_1^3(\tilde\cF_{R,\TT})  -\frac{1}{24}p_1(T\cY_6) c_1(\tilde\cF_{R,\TT})\ed
\ee
Our proposal is that, to capture all the analytic terms contributing to $\cI_{\textrm{Chiral}}$ in the limit \eqref{Cardy}, we need to evaluate the integral
\be\label{chiralprop}
\log \cI_{\textrm{Chiral}} = \int_{\cN_4\times D^2}  2\pi i  P_{\mathsf T}^{(6)}[\cB_R]\ec
\ee
where the boundary of the manifold $\cN_4\times D^2$ and metric on it is identified with the supersymmetric background \eqref{4dmetric}. 
We require that the metric on $\cN_4\times D^2$ is such that the supersymmetric Killing vector \eqref{eq:Kvector} can be extended in the bulk while retaining only a single isolated fixed point. We further demand the geometry $\cN^4\times D^2$ has a collar region near the boundary.

The rationale of \eqref{chiralprop} is as follows. Following sections \ref{redinvertible} and \ref{effact}, is that the six-form $P_{\mathsf T}^{(6)}[\cB_R]$ can be thought as an anomaly polynomial on a non-trivially fibered bulk six-manifold $\cY_6$. Integrating such polynomial along the $D^2$ fiber, with a specific choice of background fields extension, leads to a set of $3d$ Chern--Simons contact terms which can be evaluated on the reduced supersymmetric background from section \ref{susybackg} to obtain \eqref{cardychiral}. 

Here we will perform the integration on the right hand side of \eqref{chiralprop} equivariantly, following an approach inspired by \cite{Bobev:2015kza} and reviewed in appendix \ref{equivint}. Note that a similar proposal was also considered in \cite{Nahmgoong:2019hko}, our analysis from sections \ref{redinvertible} and \ref{effact} justifies why the substitution rule of \cite{Jensen:2012kj,Jensen:2013rga} appears in this problem. 

We would now like to explain why the (equivariant) reduction of the thermal anomaly polynomial reproduces the high-temperature asymptotics on the second sheet.

The basic idea is that one can formally promote the characteristic classes in \eqref{frechiralT} to equivariant characteristic classes and integrate them on $\cY_6 \equiv \bR^4_{\omega_1, \omega_2} \times \bR^2_{\mu_\TT}$ using equivariant localization. Here, the notation $\bR_{\omega_1, \omega_2}^4$ is used to specify that $U(1)\times U(1)$ acts on $\bR^4$ with equivariant parameters $\omega_1$ and $\omega_2$ identified with the superconformal index fugacities. The extra factor of $\bR^2_{\mu_\TT}$ whose $U(1)$ action has equivariant parameter denoted by $\mu_\TT$ is introduced to take into account the non-trivial disk fibration induced by the thermal circle. Thus, we can now evaluate \eqref{chiralprop} using the localization formula \eqref{local}:
\be\label{equivchiral}
\int_{\cY_6} 2\pi i P_\TT^{(6)}[\cB_R] =  {2\pi i P_\TT^{(6)}[\cB_R]|_0 \over e(T\cY_6|_0)} = {2\pi \over e(T\cY_6|_0)}\left(\frac{1}{6}c_1^3(\tilde\cF_{R,\TT})|_0  -\frac{1}{24}p_1(T\cY_6)|_0 c_1(\tilde\cF_{R,\TT})|_0\right)\ec
\ee
where $|_0$ denotes the restriction at the origin which is the unique $U(1)^3$ fixed point on $\cN_6$. Let us describe all the elements contributing to \eqref{equivchiral}:
\begin{itemize}
\item[1.)] The Euler class $e(T\cM)$ of any Riemannian manifold $\cM$ can be defined as the Pfaffian of the Riemann curvature 2-form $\cR$. From this definition we obtain
\be\nonumber
e(T\cY_6|_0) = \mu_\TT\,\omega_1\omega_2\ed
\ee
In a similar way, we can use the definition of the first Pontryagin class given in \eqref{classes} to compute
\be\nonumber
p_1(T\cY_6)|_0 = \omega^2_1 + \omega^2_2 + \mu^2_{\TT}\ec
\ee
where the last contribution comes from the  evaluation of the class $p_1(T\bR_{\mu_\TT}^2)$ as explained around equations \eqref{whitney} and \eqref{chibulk}. In our conventions for the equivariant integrals the value of the  $\mu_\TT$ will be set to $\mu_\TT^2 =(2\pi i)^2$.
\item[2.)] As explained in \cite{Bobev:2015kza}, we have to introduce an equivariant parameter for each global symmetry of the theory. Moreover, these additional parameters are identified with the corresponding flavor fugacities appearing in the index. For instance, the $U(1)_R$ equivariant parameter is identified with the $U(1)_R$ fugacity $\mu_R$ that we introduced in section \ref{susybackg}. Therefore we can compute the contribution from $c_1(\cF_R)|_0$ and get
\be\nonumber
c_1(\cF_R)^n|_0 = \mu_R^n (r-1)^n = {1\over 2^n}\,(\omega_1+\omega_2 + 2\pi n_0)^n(r-1)^n\ec
\ee
where in the last equality we used \eqref{gammaext} and $(r-1)$ is the $R$-charge of the chiral multiplet fermion.
\end{itemize} 
In view of the above analysis we finally obtain
\be
\int_{\cY_6} 2\pi i P_\TT^{(6)}[\cB_R] = {(\omega_1+\omega_2 + 2\pi i n_0)^3\over 48 \omega_1\omega_2}(r-1)^3 -  {(\omega^2_1+\omega^2_2 - 4\pi^2)(\omega_1+\omega_2+ 2\pi i n_0)\over 48 \omega_1\omega_2}(r-1)\ec
\ee
which matches the expansion $\log \cI_{\textrm{Chiral}}$ from \eqref{cardychiral} to all perturbative orders in $\beta$.
\paragraph{Generalization to Arbitrary $\cN=1$ SCFTs}
We now generalize our computation to arbitrary $\cN=1$ SCFTs, following the same general approach of the previous example.

Repeating similar steps as in the free chiral theory example, we can easily write down the thermal anomaly polynomial of an arbitrary $\cN=1$ SCFT (without additional flavor symmetries)
\be\label{fullproposal}
P_{\mathsf{T}}^{(6)}[\cB_R] = \frac{k_{RRR}}{6}c_1^3(\tilde\cF_{R,\TT})  -\frac{k_{R}}{24}p_1(T\cY_6) c_1(\tilde\cF_{R,\TT})\ec
\ee
where $k_{RRR}$ and $k_R$ are, respectively, the $U(1)_R^3$ and mixed gravitational-$U(1)_R$ anomaly coefficients. In superconformal theories, these coefficients can always be expressed in terms of the $4d$ conformal anomalies $a$ and $c$ as shown in \cite{Anselmi:1997am}: 
\be
a = {3\over 32}(3k_{RRR} -k_R)\ec \quad c = {1\over 32}(9 k_{RRR} -5k_R)\ed
\ee
As we saw above, \eqref{fullproposal} can be integrated equivariantly: 
\be\label{main}
\int_{\cY_6} 2\pi i P_\TT^{(6)}[\cB_R] = {(\omega_1+\omega_2 + 2\pi i n_0)^3\over 48 \omega_1\omega_2}k_{RRR} -  {(\omega^2_1+\omega^2_2 - 4\pi^2)(\omega_1+\omega_2+ 2\pi i n_0)\over 48 \omega_1\omega_2}k_{R}\ec
\ee
to obtain \eqref{ZCformula}. To summarize, we have established that (up to exponentially suppresed terms) the all order high-temperature expansion of the superconformal index on the second sheet is completely determined by the contribution from four-dimensional anomalies. 

Our main result \eqref{main} is very general and does not rely on a Lagrangian description of the theory. Therefore it can also be applied to strongly coupled $\cN=1$ SCFTs which would not admit a good high-temperature effective field theory description. The anomaly matching approach also allows us to verify that the $S^3\times S^1$ Casimir energy naturally contributes to the Cardy asymptotics on the second sheet. 

Our derivation can be easily generalized to other classes of $4d$ SCFTs, with different amount of supersymmetry and various flavor symmetry groups. 
It is also possible to describe superconformal theories in other (even) spacetime dimensions. The case of $2d$ SCFTs is straightforward while for 6$d$ SCFTs the story is more subtle. We will offer some preliminary remarks about such theories in section \ref{6dSCFTs}.

We would like to conclude this section by addressing the $\log |G|$ term in \eqref{ZCformula} which is associated with the one-form symmetry being spontaneously broken \cite{Cassani:2021fyv}. A $4d$ gauge theory with non-Abelian gauge group $G$ has an electric one-form symmetry $G\one = Z({G})$. As we expect the theory to be deconfined at high-temperature, these electric one-form symmetries have to be spontaneously broken \cite{Gaiotto:2014kfa, Gaiotto:2017yup}. 

If we assume that the spontaneously broken $G\one$ symmetry is minimally realized by a discrete $Z(G)$ (0-form) $3d$ gauge theory we have that
\be\nonumber
Z_{\textrm{TQFT}}[S^3] = {1\over|G|}\ec
\ee
giving the constant contribution in \eqref{ZCformula}. Note that if the latter assumption is violated, the topological theory realizing the broken one-form symmetry still has to contain the $Z(G)$ gauge theory as a subsector, and the constant term in \eqref{ZCformula} should be greater than $\log|G|$.

\section{Comments on Six-Dimensional SCFTs}\label{6dSCFTs}
In this section we discuss an application of our ideas to six-dimensional $\cN=(1,0)$ superconformal field theories. After taking into account certain assumptions that are spelled out below, we will be able to propose a conjectural Cardy formula for the $6d$ superconformal index on the second sheet. 

It is helpful to describe six-dimensional $\cN=(1,0)$ SCFTs using radial quantization. States in the Hilbert space are labeled by the energy $E$, the spins $J_1, J_2, J_3$ (denoting various diagonal combinations of $SO(6)$ Lorentz symmetry Cartans) and by the $SU(2)_R$ charge $R$. Any $(1,0)$ SCFT has both $8$ Poincar\'e and $8$ conformal supercharges. If we pick a particular supercharge $\cQ$, we can construct an Hamiltonian proportional to $\{\cQ,\cQ^\dagger\}$ and count (with a sign) BPS states annihilated by $\cQ$ \cite{Bhattacharya:2008zy,Kim:2012ava,Kim:2012qf}:
\be\label{6dindex}
\cI = \Tr_{\cH_{\textrm{BPS}}} (-1)^F e^{-\omega_1(J_1 + R)} e^{-\omega_2(J_2 + R)}e^{-\omega_2(J_3 + R)}\ed
\ee
This can be thought of as a Witten index for ${1\over 8}$-BPS local operators refined by three additional angular momentum fugacities $\omega_1, \omega_2, \omega_3$. It is also easy to introduce flavor symmetries along with their fugacities, however we choose not to do so here.

Following section \ref{CardySCFT}, we should think about the $6d$ superconfomal index $\cI$ as a holomorphic function of the fugacities $\omega_{1,2,3}$ which are allowed to take generic complex values. The index $\cI$ is not single valued function of $\omega_{1,2,3}$, it has a second sheet of validity that can be studied by shifting any of the $\omega_i$'s by $2\pi i$. For instance, let us consider the shift $\omega_1 \to \omega_1 + 2\pi i$ while keeping $\omega_2$ and $\omega_3$ fixed
\be\label{6dRindex}
\cI = \Tr_{\cH_{\textrm{BPS}}} e^{-2\pi i R} e^{-\omega_1(J_1 + R)} e^{-\omega_2(J_2 + R)}e^{-\omega_2(J_3 + R)}\ed
\ee
On the second sheet, the index counts supersymmetric states (annihilated by $\cQ$) weighted by a factor of $e^{-2\pi i R}$. Therefore, comparing with \eqref{6dindex}, we expect a completely different set of cancellations to occur in \eqref{6dRindex}. Note that most of the results from section \ref{susybackg} do not require a Lagrangian description of the supersymmetric theory. From this point of view, we assume that there exists a twisted (and squashed) $S^5 \times S^1$ supersymmetric background for the $6d$ $\cN=(1,0)$ theory which geometrize the superconformal index. Such background has not been studied in the literature and we plan to investigate its aspects in a future work. The resulting $S^5 \times S^1$ metric is characterized by three independent squashing parameters $b_{1,2,3}$ and three twisting parameters $\sigma_{1,2,3}$ which are related to the fugacities $\omega_{1,2,3}$ appearing in the index. As before, probing the second sheet corresponds to shifting one of the twisting parameters $\sigma_i$ by $2\pi i n_0$ with $n_0=\pm 1$ denoting  the first and its complex conjugate sheet.

We are now interested in exploring the asymptotic behavior of \eqref{6dRindex} in the Cardy limit 
\be\label{Cardy6d}
\omega_1,\omega_2, \omega_3 \to 0\ec ~\text{with~ ${\omega_1 \over \omega_2},{\omega_2 \over \omega_3},{\omega_3 \over \omega_1}$~fixed}\ec
\ee
using a generalization of our analysis from section \ref{proof}. Perturbative 't Hooft anomalies of $6d$ $\cN=(1,0)$ SCFTs, involving the $SU(2)_R$ symmetry and diffeomorphisms, are encoded in the 8-form anomaly polynomial \cite{Harvey:1998bx,Ohmori:2014pca, Ohmori:2014kda,Intriligator:2014eaa}:
\be\label{6danom}
P^{(8)}[\cB_R] = {1\over 4!}\Bigl(\alpha c_2(\cF_R)^2 + \beta c_2(\cF_R)p_1 + \gamma p^2_1 + \delta p_2\Bigr)\ec
\ee
where $\cB_R$ is a background gauge field for the $SU(2)_R$ and $\cF_R$ is its field strength. The anomaly coefficients
$\alpha, \beta, \gamma, \delta$ are universal and independent observables in any $\cN = (1, 0)$ SCFT. Moreover, it is also possible to express the $6d$ superconformal anomaly coefficients $a, c_1, c_2, c_3$ in terms of the 't Hooft anomaly coefficients from \eqref{6danom} using the anomaly multiplet relations \cite{Cordova:2015fha} and \cite{Beccaria:2017dmw,Yankielowicz:2017xkf} later established in \cite{Cordova:2019wns}.

As was explained in section \ref{proof}, the 8-form anomaly polynomial \eqref{6danom} is subject to the substitution rule described in sections \ref{redinvertible} and \ref{effact}. This is due to the geometric properties of the supersymmetric $S^5\times S^1$ background mentioned above. The corresponding thermal anomaly polynomial is given by
\be
P_{\TT}^{(8)}[\cB_R] = {1\over 4!}\Bigl(\alpha c_2(\tilde\cF_{R,\TT})^2 + \beta c_2(\tilde\cF_{R,\TT})p_1(T\cY_8) + \gamma (p_1(T\cY_8))^2 + \delta (p_2(T\cY_8))\Bigr)\ed
\ee
We can follow section \ref{proof} and evaluate equivariantly the above anomaly polynomial on $\cY_8 \equiv \bR_{\omega_1,\omega_2,\omega_3}^6 \times \bR_{\mu_\TT}^2$ using \eqref{local}
\be
\int_{\cY_8} 2\pi i P_{\TT}^{(8)}[\cB_R] = {2\pi P_\TT^{(8)}[\cB_R]|_0 \over e(T\cY_8|_0)}\ed
\ee
The right-hand side of the above can be further simplified using by computing the following characteristic classes
\begin{align}\nonumber
e(T\cY_8|_0) = &\,\mu_\TT\,\omega_1\omega_2\omega_3\ec \qquad p_1(T\bR_{\omega_1, \omega_2, \omega_3}^6)|_0 = \omega_1^2+\omega_2^2+\omega_3^2\ec \\ \nonumber
& p_2(T\bR_{\omega_1, \omega_2, \omega_3}^6)|_0 = \omega^2_1\omega^2_2+\omega^2_1\omega^2_3+\omega^2_2\omega^2_3\ec
\end{align}
and by setting the restrictions of $p_1(T\bR_{\mu_\TT}^2)$ and $c_2(\cF_R)$ at the origin equal to the corresponding equivariant parameters $\mu_\TT$ and $\mu_R$ as we did in section \ref{proof}. Combining all these terms together we obtain a conjectural formula for the high-temperature limit of $\cI$ on the second sheet up to exponentially suppressed terms:
\be\label{2sheet6d}
\begin{split}
\log \cI &= {1\over \omega_1\omega_2\omega_3}\left({\mu^4_R\over 24}\alpha -{\pi^2\mu^2_R\over 6}\beta + {2\pi^4\over 3}\gamma\right) + {\omega_1^2+\omega_2^2+\omega_3^2\over\omega_1\omega_2\omega_3}\left({\mu^2_R\over 24}\beta  -{\pi^2\over 3}\gamma-{\pi^2\over 6}\delta \right)\\
&+ {\omega^2_1\omega^2_2+\omega^2_1\omega^2_3+\omega^2_2\omega^2_3\over\omega_1\omega_2\omega_3}\left({\gamma \over 12}+ {\delta \over 24}\right) + {\omega_1^4+\omega_2^4+\omega_3^4\over\omega_1\omega_2\omega_3}{\gamma \over 24}\ed
\end{split}
\ee
A few comments about the above formula are in order:
\begin{itemize}
\item A more detailed analysis of \eqref{2sheet6d} requires to identify a twisted $S^5\times S^1$ supersymmetric background for the $(1,0)$ SCFT along the line of section \ref{susybackg}. We expect that, the contraction between the Killing vector and some appropriate $SU(2)_R$-symmetry background supergravity field leads (on the second sheet) to the identity
\be\nonumber
\mu_R = \half(\omega_1+\omega_2+\omega_3+2\pi i n_0)\ec
\ee
which can be plugged back in \eqref{2sheet6d} to further simplify the expression. It would also be interesting to understand if there is any positivity bound on the coefficient of the most singular terms of \eqref{2sheet6d} that could be proven from first principles.  

\item As we have seen in section \ref{CardySCFT}, determining a twisted $S^5\times S^1$ background for the superconformal index is crucial to describe a manifestly supersymmetric $5d$ high-temperature effective field theory on the second sheet. This is formulated in terms of a set of five-dimensional Chern--Simons contact terms that can be used to test our conjectural formula \eqref{2sheet6d} independently. The knowledge of such $5d$ contact terms is presently only limited to the effective theory on the first sheet, see \cite{Chang:2017cdx, Chang:2019uag}.\footnote{Despite that, see \cite{Nahmgoong:2019hko, Kantor:2019lfo} for some works in $6d$ which follow the strategy of \cite{Choi:2018hmj}.} It would be interesting to extend these results to the second sheet and obtain the complete set of $5d$ supersymmetric Chern--Simons terms describing the high-temperature expansion.
\end{itemize}

\section*{Acknowledgments}\noindent We thank Pieter Bomans, Davide Cassani, Christian Copetti, Gabriel Cuomo, Justin Kaidi, Zohar Komargodski, Alba Grassi and Yuji Tachikawa for helpful discussions. The work of K.O.\ has been supported by the Simons Collaboration on Global Categorical Symmetries. L.T.\ is supported by IISN-Belgium (convention 4.4503.15). Any opinions, findings, and conclusions or recommendations expressed in this material are
those of the authors and do not necessarily reflect the views of the funding agencies.

\bigskip
\appendix
\section{Equivariant Integration}\label{equivint}

In this section we introduce some mathematical concepts that are quite useful when dealing with the integration of the anomaly polynomial. 

Consider a smooth $2\ell$-dimensional manifold $\cM_{2\ell}$ with a $T= U(1)^n$-action. For such class of manifolds one can define a cohomology theory known as equivariant cohomology, denoted by $H^{\bullet}_T(\cM)$. In analogy with ordinary cohomology theory one can also introduce the concept of equivariant differential forms and cohomology classes. We refer the interested reader to \cite{Cordes:1994fc, Blau:1995rs, 2007arXiv0709.3615L} for detailed expositions of this subject. Assuming that there is only a discrete set of isolated fixed points of the $T$-action on $\cM$, we can invoke the powerful  Atiyah--Bott--Berline--Vergne localization formula for an equivariant cohomology class $[\alpha] \in H^{\bullet}_T(\cM)$:
\be\label{local}
\int_{\cM_{2\ell}} \alpha = \sum_{p} {\alpha|_p \over e(T\cM|_p)} \ec
\ee
where $p$ runs over all the T-fixed points, $\alpha|_p$ and $T\cM|_p$ are restrictions of $\alpha$ and $T\cM$ at $p$. See \cite{Cremonesi:2013twh} for a particularly clear review. 

It was pointed out in \cite{Bobev:2015kza} that the anomaly polynomial for theories on a space with $T$ action can be extended equivariantly by promoting all the characteristic classes to equivariant characteristic classes. The anomaly polynomial can be thus integrated equivariantly using the Atiyah--Bott--Berline--Vergne localization formula. Furthermore, to evaluate the right hand side of \eqref{local} one also identifies the formal equivariant parameters for the $T$ action with flavor symmetry fugacities of the original theory. This idea has had numerous application in the supersymmetric context, see for example \cite{Alday:2009qq, Nahmgoong:2019hko, Bah:2019rgq, Hosseini:2020vgl}.

Note that, in equivariant cohomology, it is standard to have equivariant forms whose degrees are greater than the dimension of $\cM$ and whose equivariant integrals are non-vanishing. In this sense, the reduction along the fiber in \eqref{termalmatch} is completely natural from the point of view of equivariant integration \cite{Cordes:1994fc, Blau:1995rs}.

Let us discuss how this procedure can be carried out in a simple example without supersymmetry. We would like to study a $2d$ theory on a thermal background $\cM_2$ with metric \eqref{KKmetric}
\be\label{KK2d}
ds^2 = (d\tau + a(x)dx)^2 + dx^2\ec
\ee
where $\tau \simeq \tau +\beta$. As in the four-dimensional case, we can introduce a background gauge field for the $U(1)_{\TT}$ isometry $\cA_\TT$ and study the high-temperature limit. In what follows we will focus on the following contribution to the effective action \cite{Banerjee:2012iz}
\be\label{0der}
\cW^{[0]}_{\QFT}[\cA_\TT]= {2\pi k_{2d} \over \beta} \int a \ec
\ee
and show how to fix the coefficient $k_{2d}$ via equivariant integration. The thermal anomaly polynomial in this case should be obtained by the $2d$ gravitational anomaly polynomial by
\be\label{p4t}
P_{\TT}^{(4)} = {k_{\cR^2}\over 24}p_1(T\cY_4)\ed
\ee
where $\cY_4$ is obtained by extending the metric \eqref{KK2d} as explained in the main text. We work with a convention where $k_{\cR^2}= 1$ for a single left-moving Weyl fermion.

As explained in \cite{Bobev:2015kza}, we now need to extend $P_{\TT}^{(4)}$ equivariantly and compute: 
\be
\int_{\cY_4} P_{\TT}^{(4)}=  \sum_{p} {P_{\TT}^{(4)}|_p \over e(T\cM|_p)}\ed
\ee
with $\cY_4= \bR^2_{\beta}\times \bR_{\mu_{\TT}}^2$. Here  $\bR_{\beta}^{2}$ denotes $\bR^2$ endowed with an equivariant parameter which we identify with the radius of the thermal circle $\beta$. Such space features prominently in the study of $2d$ supersymmetric gauge theories \cite{Nekrasov:2009rc, Nekrasov:2010ka}.  Since the origin of $\cY_4$ is the unique fixed point under the action of $U(1)\times U(1)$, we obtain
\be
\int_{\cY_4} 2\pi iP_{\TT}^{(4)}=  {2\pi ik_{\cR^2}\over 24}{p_1(T\cY_4)|_0\over e(T\cY_4|_0)} =  {2\pi ik_{\cR^2}\over 24}{\mu_\TT^2\over e(T\cY_4|_0)}\ec
\ee
where $\mu_\TT$ is the fugacity for the $U(1)_\TT$ symmetry which in our conventions is set to $\mu_\TT = 2\pi i$. Substituting the value of $\mu_{TT}$ and comparing with the coefficient of \eqref{0der}, we conclude that
\be
k_{2d} = -{\pi k_{\cR^2}\over 12}\ed
\ee
This agrees with the result obtained in \cite{Jensen:2012kj, Jensen:2013rga, Golkar:2015oxw}.

\makeatletter
\interlinepenalty=10000
\bibliographystyle{utphys.bst}
\bibliography{CardySubs.bib}
\makeatother

\end{document}